\documentclass[letterpaper,11pt]{article}
\usepackage{jheppub}
\usepackage{amsmath,latexsym}
\usepackage{booktabs} 
\usepackage{graphicx} 
\usepackage{siunitx}
\usepackage{setspace}
\usepackage{amssymb}
\usepackage{amsmath}
\usepackage{amstext}
\usepackage{amsfonts}
\usepackage{bbold}
\usepackage{slashed}
\usepackage{comment} 
\usepackage{tikz-feynman}
\usepackage{xcolor}
\usepackage{subcaption}
\usepackage{physics}
\usepackage{slashed}
\usepackage{graphics}
\usepackage{setspace}
\usepackage{amssymb}
\usepackage{amsmath}
\usepackage{amstext}
\usepackage{amsfonts}
\usepackage{bbold}
\usepackage{slashed}
\usepackage{tikz-feynman}
\usepackage{xcolor}
\usepackage{subcaption}
\usepackage{physics}
\usepackage{slashed}
\newcommand{\beq}{\begin{equation}}
\newcommand{\eeq}{\end{equation}}
\renewcommand{\l}{\left}
\renewcommand{\r}{\right}

\newcommand{\bea}{\begin{eqnarray}}
\newcommand{\eea}{\end{eqnarray}}

\newcommand{\nn}{\nonumber}

\newcommand{\be}{\begin{eqnarray}}
\newcommand{\ee}{\end{eqnarray}}

\definecolor{SGreen}{rgb}{0.0547,0.613,0.328}
\definecolor{NodeBlue}{rgb}{0.0547,0.148,0.578}

\begin{document}

\title{On the Motion of Compact Objects in Relativistic Viscous Fluids} 

\author{ Beka Modrekiladze, Ira Z. Rothstein and  Jordan Wilson-Gerow}
\affiliation{Department of Physics, Carnegie Mellon University, Pittsburgh, PA 15213, USA}

\abstract{

We present a world-line  effective field theory of compact objects moving relativistically through a viscous fluid.
The theory is valid  when velocity gradients are small compared to the inverse size of the object. Working within  the EFT eliminates the need to solve a boundary value problem by turning all interactions between the fluid and the object into a source term  in the action. We use the EFT to derive the relativistic equations of motion for a compact object immersed in a viscous fluid in a curved background.

}

\maketitle

\section{Introduction}

Solving the Navier-Stokes equations presents a significant challenge which, for realistic three dimensional cases, at present, can only be attacked numerically. The inclusion of immersed 
 dynamical objects considerably compounds this challenge
 as do relativistic and gravitational effects that
 arise, e.g., when simulating binary inspirals.
The most severe challenge presented by the addition of dynamical objects is a set of  
boundary conditions that must be updated at each time step. 
 To quote from the review on Smoothed Particle Hydrodynamics (SPH) \cite{SHADLOO201611}, "\textit{the treatment of boundary conditions is certainly one of the most difficult technical points of the SPH method, so that it still remains a very active research topic}." This challenge is even more pronounced in grid-based simulations.  
This paper aims to introduce a formalism that will mitigate the computational hurdles presented by this problem, all  within a first
principles approximation. Our formalism also facilitates analytic calculations when non-linearities can be treated as perturbations.

Let us start with the simple case of  an infinitely stiff spherical body moving through a viscous fluid.
We will expand the ratio $R/r$, with $r$ being the typical scale for fluid gradients and $R$ being the object's size. In this limit it is natural to work within the confines of a Point Particle Effective Field Theory (PPEFT)
which  has been utilized to great success in  various contexts including gravitational wave
production from binary inspirals \cite{Goldberger:2004jt,Goldberger:2006bd}, the interactions between defects on fluctuating membranes \cite{Rothstein:2011ku} and Casimir forces on cogs \cite{Vaidya:2013dza} and monopole catalysis \cite{Bogojevic:2024xtx}. In these cases, the EFT is used to
 systematically  generate analytic solutions from first principles. However, in the present context, an EFT will be used to
 simplify numerical approximations which in some cases may be intractable otherwise.
It will also allows to write down relativistic equations of motion for compact objects in a viscous fluid.

 The EFT  
 starts with a world-line action for a point particle, and corrections due to finite size effects 
 are accounted for by adding higher dimensional terms. These terms are accompanied by coefficients whose
 values encapsulate the short distance physics which has been coarse grained. In the context of our problem
 of interest here, the power of the EFT is manifested by the fact that we can effectively eliminate the need to impose
 boundary conditions at each time step. All of  boundary effects in the EFT are captured by the introduction
 of an effective coupling in the action, whose value can be fixed by a matching calculation, as will be discussed below, 
 which can, in general, be done analytically since the matching coefficient can be extracted for a fixed (symmetric) fluid
 configuration.

 Since we are interested in immersing the particle in a fluid, the full EFT will include a bulk action for the fluid which
 will couple to the point particle.
 We will also show how viscosity affects  the motion of the body
 in a generally covariant way at the level of the action using the in-in formalism.
 The bulk action in the in-in formalism was presented in \cite{Liu:2018kfw} \footnote{For other approaches to relativistic viscous hydrodynamics \cite{Bemfica:2020zjp}.}, and we will generalize this result to allow for viscous interactions between the body and the fluid. This will allow us to write down a generally covariant equation of motion for a body in a viscous fluid which, to our knowledge, has not appeared in the literature\footnote{Although the form of the equation was clearly anticipated by \cite{Correia:2022gcs}}.

 \section{The Fluid EFT}

The Lagrangian approach\footnote{There is a long history of Lagrangian approaches to the field theory of fluids, for references see \cite{Soper:1976bb}. For a more modern discussion see \cite{Dubovsky:2005xd,Andersson:2006nr}.} to fluid dynamics is usually based upon one of two variable choices which can be interpreted as two distinct frames.
In the Eulerian variables $\phi^I(x,t)$ can be thought of as labeling the fluid element that is sitting at $x$ and time $t$, while the Lagrangian frame describes
the physics in terms of $X^I(\phi^I,t)$ which gives the position at time $t$ of the particle whose label is given by $\phi^I$. In this paper we will be working in the Eulerian frame where
the comoving volume form  is given by
\beq
\Omega (\phi)= \frac{1}{3!}\epsilon^{IJK} d \phi^I\wedge d\phi^J\wedge d \phi^K 
\eeq
We can build an algebraically conserved one form current 
\beq
J=*\Omega,
\eeq
and in the non-relativistic limit we may associate $d*J=0$ with the  conservation of particle number\footnote{For  relativistic fluids the conservation law becomes that of the entropy current.}.
$J^2$ is the square of the (rest frame) density and can be written as $J^2=\rho^2\equiv  \det(B_{IJ})$ where
$B_{IJ}=\partial_\mu \phi^I \partial^\mu \phi^J$. The fluid four velocity  $u^\mu$ is defined as a vector field which preserves the value of the label $\phi^I$, i.e.  $u \cdot \partial \phi^I=0$, and is normalized so that in the fluids local rest frame
$u^\mu=(1,\vec 0)$, i.e. $u^\mu= J^\mu/\sqrt{J^2}$ is the dimensionless four velocity normalized to one and we work in units where $c=1$.

A fluid is defined in terms of an action which is  invariant under volume preserving
spatial diffeomorphisms, 
\beq
\phi^I \rightarrow  \phi^{\prime I}
\eeq
with
\beq
 \left| \textrm{Det} \frac{\partial \phi^{\prime I}}{\partial \phi^J} \right|  =1.
\eeq
This symmetry defines a fluid. The power counting is such that one derivative on the field
is leading order and subsequent derivatives are suppressed by powers of the UV cut-off, which in this hydrodynamic approximation
is the mean-free-path.  The leading order action is written in the convenient form 
\beq
S= \int d^4x\, \rho U(\rho^{-1}),
\eeq
where $U$ is some unknown function.

 Typically in an EFT we would expand in some background solution allowing us to 
generate a polynomial action with a set of Wilson coefficients that need to be fixed by matching. But when working
in hydrodynamics we are interested in the Cauchy problem, not just small oscillations around a ground state. Furthermore, it does
not behoove us to calculate the Euler-Lagrange equations in terms of the Euler variables $\phi^I$. Instead we use the fact the hydrodynamic equations
of motion are equivalent to stress energy conservation
where the  stress tensor follows from the action
\beq
T^{\mu \nu}=  u^\mu u^\nu(\rho U-U^\prime) +g^{\mu \nu}U^\prime.
\eeq
Considering the fluid at rest allows us to identify $\rho U$ with the energy density  ${\cal E} $. Moreover, comparison with a perfect fluid form
\beq\label{fluidST}
T^{\mu \nu}= u^\mu u^\nu({\cal E}+P)-g^{\mu \nu}P
\eeq
allows us identify the pressure as well, $U^\prime=-P$ and $ U=\frac{{\cal E}}{\rho}$.

 \section{The Point Particle EFT in a  Perfect Fluid}
 
 Typically world-line effective  theories  consider the dynamics of an object around some fixed background with a set of isometries.
 Here we wish to push the formalism a little further by considering the
 case where the object moves in a time-evolving background, i.e. a fluid flow.   
 
 In the world-line approximation we assume that the gradients in the fluid are small compared to the inverse size of the
 particle.  Conservatively the EFT will be under best control for laminar flow/low Reynolds number where the velocity gradients near the surface of the
 body are relatively small. This would seem to pose a challenge to cases where the fluid boundary conditions are such that the velocity of the fluid is discontinuous.
 However, since the viscosity ($\eta$)\footnote{For a perfect fluid the generation/non-generation of a boundary layer leads to D'Alembert's paradox discussed below. } smooths out abrupt changes in velocity,
  we would expect that the velocity gradients to scale as
 \beq
 \frac{dv}{dr} \sim \frac{1}{\eta}.
 \eeq
 We will also consider the case 
 of potential flow of a perfect fluid, under situations where  the flow is laminar. Away from the body, where the EFT is valid, all gradients die off with higher powers of $1/r$.

To make our theory consistent with covariance, the
 object must be deformable. We could decide to treat the deformations as new world-line degrees of freedom, or if we assume that
 the time scale for the changes in the fluid velocity field is large compared to signal crossing time in the interior, we can
 account for the effect of these changes by including terms in the action which describe the, effectively instantaneous, response due the changing shape of the body.  In the case of the gravitational world-line theory such terms are tidal response operators \cite{Goldberger:2005cd}  proportional to the Weyl tensor squared  and are
suppressed by derivatives of the gravitational field. In the present context such finite size effects will be down by higher order gradients of the fluid field.

 To build the EFT we parameterize the particle's world-line via $x^\mu(\lambda)$.
 The relevant symmetries are: world-line reparameterization, Lorentz invariance and volume preserving diffeomorphisms acting 
 on the fluid field,   $\phi^I$. The most general action which satisfies these symmetries, and is leading order in the derivative expansion, is of the form 
 \beq\label{eq:EFTaction}
 S=-M\int d\lambda\,\sqrt{\dot x^2}   F\l( \frac{\dot x \cdot u}{\sqrt{\dot x^2}}, \rho\r)
 \eeq
 where $F$ is a dimensionless unknown function, and $M$ is the object's mass. 
 We notice that the combination
 \bea
 \label{gamma}
 \gamma =  \frac{\dot x \cdot u}{\sqrt{\dot x^2}}= 1+ (\vec u-  \dot {\vec x})^2 f(\vec u, \dot{\vec x}).
\eea
 is approximately one when the relative velocities vanish, and we will use this fact
 to expand $F$ around this kinematic point.

 From this action, we can extract the stress-energy tensor, 
\begin{align}\label{eq:ST}
    T^{\mu\nu}&=M\int d\lambda\,\frac{\delta^{4}(x-x(\lambda))}{\sqrt{-g}}\bigg[\frac{\dot x^{\mu} \dot x^{\nu}}{\sqrt{\dot x^2}}F(\gamma, \rho) +\sqrt{\dot{x}^{2}}(u^{\mu}u^{\nu}-g^{\mu\nu}) \rho \frac{\partial F(\gamma, \rho)}{\partial  \rho} \nn \\
    &+\sqrt{\dot{x}^{2}}\frac{\partial F(\gamma,\rho)}{\partial \gamma} \left(\frac{\dot x^{\mu}u^{\nu}+u^{\mu}\dot x^{\nu}}{\sqrt{\dot x^{2}}}-\gamma u^{\mu}u^{\nu}-\gamma\frac{\dot x^{\mu}\dot x^{\nu}}{\dot x ^{2}}\right)\bigg]\,.
\end{align}

 Note that for the case of the fluid we were able to match for an unknown function ($U$) using covariance (\ref{fluidST}) and the known form 
 for the stress energy tensor for a fluid. We could try to follow the same process in this case
 but now we have four tensor structures (as opposed to two in the 
 case of the fluid) and even if we assume an internal equilibrium it's not clear how to match these functions. To gain predictive power
 we will expand the function $F$ around $\gamma=1$.\footnote{This is similar to what  happens when considering the motion of a particle in Einstein-Aether theory \cite{Foster:2007gr}.} 
 In light of eq.(\ref{gamma}), this expansion corresponds to one in small relative velocities
 which we may consider as a generic configuration. Morever this limit overlaps with   
   the non-relativistic limit where we will be performing our matching. In general, we will put our results in relativistic form.

In the non-relativistic limit the fluid's particle density will be approximately proportional to its energy density,
\begin{equation}
    \mathcal{E}(\rho) = \mu \rho + \mathcal{O}(c^{-2})\,,
\end{equation}
where $\mu$ is the mass of the underlying fluid particles. The object will respond primarily to the mass/energy density and current of the fluid, as opposed to the particle density. We then introduce the appropriate dimensionless variable, 
\bea
\hat \rho = \frac{\rho \mu}{\rho_{\textrm ob}}\,,
\eea
where $\rho_{\textrm ob}$ is the density of the object.
$\hat \rho$ sets the magnitude the  of drag forces. Without loss of generality we can then consider our action to be a function of $\hat{\rho}$ rather than $\rho$.
 

 The relativistic stress energy tensor then takes the form
\begin{align}\label{ST}
    T^{\mu\nu} =& M \delta^{3}(x-x(t))\bigg[\frac{\dot x^{\mu} \dot x^{\nu}}{\sqrt{\dot x ^{2}}}\left(F(1,\hat{\rho})-F'(1,\hat{\rho})\right)+\left(\dot{x}^{\mu}u^{\nu}+u^{\mu}\dot{x}^{\nu}-\gamma u^{\mu}u^{\nu}\sqrt{\dot x^{2}}\right)F'(1,\hat{\rho}) \nn \\
    &+\sqrt{\dot x ^{2}}\left(u^{\mu}u^{\nu}-g^{\mu\nu}\right)\left(\hat{\rho} \frac{\partial F(1,\hat{\rho})}{\partial \hat{\rho}}+(\gamma-1)\hat{\rho}\frac{\partial F'(1,\hat{\rho})}{\partial \hat{\rho}}\right)\bigg]+\mathcal{O}((\gamma-1)^2)\,,
\end{align}
where prime denotes derivative with respect to the first argument. The four-velocities in the lab frame are given by
\begin{equation}
    \dot{x}^{\mu} = \frac{(1,\vec{v})}{(1-\vec{v}^{2})^{1/2}} \qquad \textrm{and}\qquad u^{\mu} = \frac{(1,\vec{u})}{(1-\vec{u}^{2})^{1/2}}\qquad \textrm{and}\qquad \gamma=\frac{1-\vec{u}\cdot\vec{v}}{(1-\vec{u}^{2})^{1/2}(1-\vec{v}^{2})^{1/2}}
\end{equation}

 In this limit the action is then given as 
 \beq
 S=-M\int d\lambda  \sqrt{\dot x^2}\l(F(1,\hat{\rho})+ F^\prime\l( 1, \hat{\rho}\r)(\gamma-1)+\mathcal{O}((1-\gamma)^2)\r)
 \eeq

 \subsection{Matching}\label{sec:matching}

To match  $F(1,\hat{\rho})$ and $F^\prime(1,\hat{\rho})$ we consider equating $T_{00}$ in the full and effective theory.
We are free to match in the non-relativistic limit since Lorentz symmetry will allow us to lift the result to the full relativistic action.
 Consider $T_{00}$ in the static limit, $u=v=0$
 \bea
\int_V T^{EFT}_{00}&=&M_{fl}+MF(1,\hat \rho) \nn \\
\int_V  T^{full}_{00}&=&M_{fl}-\mu\rho V_p +M.
 \eea
 where $M_{fl}$ is the total mass of the fluid and $V_p$ is the volume of the ball. 
  The internal energy of the sphere has been absorbed into the mass. Crucially, in the full theory the fluid energy density is integrated only outside the ball, producing the $-\mu\rho V_{p}$ contribution, while in the EFT the ball has been treated as a point particle and the fluid energy density is integrated over all of space. The overcounting of fluid density in the EFT is precisely accounted for by the matching coefficient,
 \beq
 \label{F}
 \boxed{ F=1-\hat \rho.}
 \eeq
 

 To match $F^\prime$ we calculate the force on the particle in both full and effective theory,
 \beq
 \label{eom}
 \frac{d}{dt} \frac{\partial L}{\partial v^i}=  M((1-\hat \rho)\dot v^i+F^\prime (\dot u^i- \dot v^i))=MF^\prime(\vec v-\vec u)\cdot\partial^i \vec u
 \eeq
 where we have used eq. (\ref{F}) and the incompressibility condition.
 In the full theory we can solve the problem for a potential flow around a sphere and the result for the force on  a sphere
 is given by (see e.g. \cite{landau1959fm})
 \beq
 \label{landau}
  \dot v^i=\frac{1}{2}\hat \rho(3\dot u^i - \dot v^i).
 \eeq
  We are immediately struck by the fact that in the full theory there is no net force on the particle in a constant flow. This is known as d'Alembert's paradox \footnote{Perhaps it would more appropriately called a ``puzzle''. What is even more interesting phenomena is that while the drag force becomes finite when viscosity is turned on, it does not drop to zero as the  viscosity is taken to zero. This is known as the ``dissipative anomaly" (see e.g. \cite{RevModPhys.78.87}).}. 
  
 There seems to be an inconsistency between the full and the effective theory. But we need to take care when interpreting the
 equations of motion. In the EFT we treat the delta function source as perturbation, so we write $\vec u= \vec u_0 +\delta \vec u$, 
 for some constant background $\vec u_{0}$. Then to leading order we can replace $\vec u$ by $\vec u_0$ in the RHS of equation (\ref{eom}),  which then vanishes in agreement with the full theory result. 
Matching coefficients then yields
  \beq
 \label{Fp}
 \boxed{ F^\prime=-\frac{3}{2}\hat \rho.}
 \eeq

\subsubsection{Solving the Modified Euler Equation: Alternative Matching Method}

As previously stated, the power of the EFT is that it allows us to morph a dynamical  boundary value problem 
into a problem with no boundaries but with localized sources on the right hand side of the fluid equations. In the full fluid theory the fluid equations of motion are equivalent to energy-momentum conservation. In the effective theory, provided that the particle also satisfies its equation of motion, the fluid equations of motion are equivalent to conservation of \emph{total} energy-momentum conservation. The non-zero energy momentum of the particle then appears as localized sources in the continuity and Euler equations.

As another check on our previous matching we will see how the EFT reproduces the full theory fluid flow for a body at rest in a fluid which asymptotically is constant and non-zero in the z-direction in the non-relativisitic limit. We will leave $F(1,\hat{\rho})$ and $F'(1,\hat{\rho})$ unfixed in the subsequent calculation, and determine their values by matching the fluid flow. Using the fluid and particle energy momentum tensors, eqs.(\ref{fluidST}) and (\ref{ST}), we obtain the continuity equation
\bea
\partial_{\mu}T^{\mu 0}=\partial_{0}\left(\rho_{ob}\hat\rho+\delta^{3}(x)MF\right)+\partial_{j}\left(u^{j}\rho_{ob}\hat\rho+u^{j}\delta^{3}(x)M\hat{\rho}\frac{\partial F}{\partial \hat{\rho}}\right)=0\,.
\eea
For static and incompressible flows this reduces to 
\beq\label{eq:continuity}
\partial_{j}u^{j}=-V_{p}\frac{\partial F}{\partial \hat{\rho}}\partial_{j}\left(u^{j}\delta^{3}(x)\right),
\eeq
where $V_{p}=M/\rho_{ob}$ is  the volume of the body.

Conservation of momentum gives the Euler equation 
\bea\label{eq:eulereqn}
\partial_{\mu}T^{\mu i} &=& \partial_{0}\left(u^{i}\left(\rho_{ob}\hat{\rho}+\delta^{3}(x)M\hat{\rho}\frac{\partial F}{\partial \hat{\rho}}\right)\right) +\partial_{j}\delta^{ij}\left(P+\delta^{3}(x)M\hat{\rho}\frac{\partial F}{\partial \hat{\rho}}+\frac{u^{2}}{2}\delta^{3}(x)M\hat{\rho}\frac{\partial F}{\partial \hat{\rho}}\right) \nn \\
&+&\partial_{j}\left(u^{i}u^{j}\left(\rho_{ob}\hat{\rho}+\delta^{3}(x)M\hat{\rho}\frac{\partial F'}{\partial \hat{\rho}}-\delta^{3}(x)MF'\right)\right)\,.
\eea

Before proceeding to solving these equations are finite fluid velocity, let's look at the simplest case, $u^{i}=0$. Here there is already a small lesson to be learned. There is an exact solution which has $u^{i}=0$, for which the Euler equation only requires that the total stress is homogeneous in space
\beq
T^{ij}=\delta^{ij}\left(P_{static}+\delta^{3}(x)M\hat{\rho}\frac{\partial F}{\partial \hat{\rho}}\right) = \delta^{ij}P_{0}\,.
\eeq
The Euler equation dictates that fluid pressure $P_{static}$ is the sum of a constant term and a delta function term.  The singular term serves only to cancel out the singular stress on the worldline, ensuring that the \emph{total} stress is constant throughout space. Once we turn on finite fluid velocity, we will again find that both the stress on the particle and the solution for the fluid pressure contain distributional singularities but in the total stress these two contributions cancel out, as they must.

Let us now look for static and incompressible solutions to these equations which have finite fluid velocity.  To do so, we will treat the particles effects on the flow perturbatively and expand the fluid about the constant solution $u_{j}=U^{j}+\delta u^{j}$, $P=P_{0}+\delta P$.

 In doing this expansion, we must recognize that terms of the form $\delta^{3}(x)\delta u^{j}(x)$ are singular since $\delta u^{j}(x)$ is itself sourced by the delta function.  These singularities are well understood in the context of effective field theory and are readily accounted for by suitable regularization and renormalization of the parameters of the theory.  For our purposes, we will implicitly work in an analytic regularization scheme, e.g. dimensional regularization, in which scaleless divergences like $\delta^{3}(x)\delta u^{j}(x)$ are automatically renormalized to zero.  With this understood, the linearized Euler equation is simply 
\bea\label{eq:linearizedEuler}
\rho_{ob}\hat{\rho}U^{j}\partial_{j}\delta u^{i}+\partial^{i}\delta P = MF' U^{i}U^{j}\partial_{j}\delta^{3}(x)-\frac{1}{2}U^{2}M\hat{\rho}\frac{\partial F'}{\partial \hat{\rho}}\partial^{i}\delta^{3}(x)-M\hat{\rho}\frac{\partial F}{\partial \hat{\rho}}\partial^{i}\delta^{3}(x)\,.
\eea

By contracting the Euler equation with the background velocity $U^{i}$, we can identify a Bernoulli equation which specifies a quantity that is constant along the background streamlines, i.e. along the $U^{i}$ direction. Since we require the perturbations $\delta P, \delta u^{i}$ to decay asymptotically, this Bernoulli equation gets lifted to a quantity which is constant over all of space, giving an expression for the pressure
\beq\label{eq:deltaP}
\delta P = -\rho_{ob}\hat{\rho}U_{j}\delta u^{j} + \delta^{3}(x)M\left(U^{2}F'-\frac{U^{2}}{2}\hat{\rho}\frac{\partial F'}{\partial \hat{\rho}}-\hat{\rho}\frac{\partial F}{\partial \hat{\rho}}\right)\,.
\eeq
Inserting this back into \eqref{eq:linearizedEuler} we obtain
\beq
U^{j}(\partial_{j}\delta u_{i}-\partial_{i}\delta u_{j})=\frac{V_{p}F'}{\hat{\rho}}U^{j}(U_{i}\partial_{j}-U_{j}\partial_{i})\delta^{3}(x)\,.
\eeq
Note that this only determines the vorticial part of the flow, $\omega^{i}= \epsilon^{ijk}\partial_{j}u_{k}$, and we'll need to check the continuity equation afterwards to determine whether additional longitudinal parts of $u_{j}$ need to be added. 

In terms of the vorticity the linearized Euler equation takes the simple form,
\beq
\epsilon_{ijk}U^{j}\omega^{k}=\frac{V_{p}F'}{\hat{\rho}}U^{j}U_{l}\epsilon_{ijk}\epsilon^{klm}\partial_{m}\delta^{3}(x)\,.
\eeq
This determines the vorticity up to terms tangent to  $U^{j}$.  However, since there are no other tensors that we can construct which are odd under parity and point along the $U^{j}$ direction we can immediately solve for the vorticity
\beq
\omega^{k}=\frac{V_{p}F'}{\hat{\rho}}\epsilon^{klm}U_{l}\partial_{m}\delta^{3}(x)\,,
\eeq
We see that the object is acting as a dipolar point source for vorticity. This is quite analogous to the point source of a magnetic dipole, i.e. an infinitesimal current loop. The solution is still effectively potential flow away from the source since there is no ``monopole'' source for the vorticity. 
Using the standard Helmholtz decomposition of the flow into the sum of a gradient term and the curl of a vector potential, we can derive the Biot-Savart-like integral solution
\bea
\delta u^{i} &=& -\epsilon^{ijk}\partial_{j}\nabla^{-2}\omega_{k} \nn\\
&=&-\frac{V_{p}F'}{\hat{\rho}}\frac{1}{4\pi}\left(\frac{U^{i}}{|x|^{3}}-\frac{3(U^{j}x_{j})x^{i}}{|x|^{5}}\right)\,.
\eea
In writing this we've discarded a term proportional to $\delta^{3}(x)$. Since the solutions to the equation of motion in the EFT are only valid as an expansion in powers of $V_{p}/|x|^{3}$, the delta function term is an unphysical artifact. Its appearance here is not surprising (see e.g. \cite{Jackson:1998nia}), but since we've already dropped other short distance contributions when we performed the perturbative expansion to solve the equation of motion, to remain consistent we must also discard the delta function term. 

The solution for potential flow around a static sphere of radius $R$ is elementary and is given by, see e.g. \cite{landau1959fm} \, 
\beq    
u^{i}=U^i + \frac{R^{3}}{2}\left(\frac{U^{i}}{|x|^{3}}-\frac{3(U^{j}x_{j})x^{i}}{|x|^{5}}\right).
\eeq
Matching the EFT to this exact solution we determine
 \beq\label{eq:Fpmatched}  \boxed{
F^\prime=-\frac{3}{2} \hat{\rho}}\,,
\eeq
 in agreement with \eqref{Fp} which had been derived by an equation of motion argument.

As previously mentioned, since the Euler equation only determined the vorticity we need to also check that our solution is consistent with the continuity equation \eqref{eq:continuity}.   The careful statement of \eqref{eq:continuity} is obtained if we multiply by $x^{i}$ integrate over some region $\mathcal{B}$, and integrate by parts to lift the derivative off the delta function, that is
\beq\label{eq:contcheck}
\int_{\partial \mathcal{B}}d^{2}x\, n_{j}x^{i}\delta u^{j} -\int_{\mathcal{B}}d^{3}x\, \delta u^{i} = V_{ob}\frac{\partial F}{\partial \hat{\rho}}U^{i}
\eeq
It is straightforward to verify that the integral of the dipole solution $\delta u^{i}$ over a sphere is zero, and so the bulk integral in \eqref{eq:contcheck} vanishes. Performing the surface integral over a sphere with constant radius, we find
\beq\label{eq:contcheck2}
\int_{\partial \mathcal{B}}d^{2}x\, n_{j}x^{i}\delta u^{j} = \frac{2}{3}\frac{MF'}{\rho_{ob}\hat\rho}U^{i}\,.
\eeq
Consistency with the continuity equation then requires
\begin{equation}
    F'=\frac{3}{2}\hat{\rho}\frac{\partial F}{\partial \hat{\rho}}\,,
\end{equation}
which implies via \eqref{eq:Fpmatched} that
\begin{equation}
\frac{\partial F}{\partial \hat{\rho}}=-1\,,
\end{equation}
in  agreement with \eqref{F}.

As another check we can compute the total pressure, i.e. the isotropic part of the stress tensor
\begin{equation}
    T^{ij}\supset \delta^{ij}\left(P+\delta^{3}(x)M\hat{\rho}\frac{\partial F}{\partial \hat{\rho}}+\frac{u^{2}}{2}\delta^{3}(x)M\hat{\rho}\frac{\partial F}{\partial \hat{\rho}}\right)\,.
\end{equation}
Linearizing, and inserting our solutions for $\delta P$ and $\delta u_{i}$ we find the total pressure
\begin{equation}
    T^{ij}\supset \delta^{ij}\left(P_{0}-\frac{\mu\rho\,U^{2}}{2}\frac{R^{3}}{r^{3}}\left(1-3\cos^{2}\theta\right)\right)\,,
\end{equation}
in  agreement with the $\mathcal{O}((R/r)^{3})$ pressure known for the exact solution. 

\subsection{Incompressibility and the Point Particle Limit }

One might still wonder how it is that we could assume that the fluid density was constant throughout all of space, given that the density should be zero ``inside the object''.  As is standard for the incompressible limit, the five unknown quantities $u^{i}, P, \rho$ are reduced to four by prescribing the profile of the mass density and using the four equations coming from energy-momentum conservation to determine $u^{i}$ and $ P$. In the full theory a natural choice is that the density be constant. In the EFT we might expect that the density is constant everywhere except on the location of the particle.  An ansatz for  the functional form of the density which models this is 
\begin{equation}\label{eq:tempdensity}
\mu\rho(x) = \mu\rho_{0} -\alpha\mu\rho_{0}V_{p}\delta^{3}(x)\,,
\end{equation}
with some parameter $\alpha$ to account for the missing fluid.  This is not, however, what we did when solving the equations in the previous section, where we simply took the density as a constant. Our assumption that $\alpha=0$ may seem further invalidated by the fact that the pressure solved for in \eqref{eq:deltaP} contained a singular piece at the location of the object. 

The resolution of this confusion is the fact that the part of the density profile \eqref{eq:tempdensity} which is localized to the worldline can always be absorbed into a redefinition of the matching function $F(\gamma,\rho)$. To see this particularly clearly, notice that the energy momentum tensor \eqref{eq:ST} contains a contribution of the form
\begin{align}\label{Tpp}
    T^{\mu\nu}&\supset M\int d\lambda\,\frac{\delta^{4}(x-x(\lambda))}{\sqrt{-g}}\bigg[\sqrt{\dot{x}^{2}}(u^{\mu}u^{\nu}-g^{\mu\nu})\hat \rho \frac{\partial F(\gamma,\hat \rho)}{\partial \hat \rho} \nn \\
    &-\gamma\sqrt{\dot{x}^{2}}\frac{\partial F(\gamma,\hat \rho)}{\partial \gamma}  u^{\mu}u^{\nu}\bigg]\,.
\end{align}
These tensor structures are exactly those which appear in the fluid stress tensor \eqref{fluidST}, so we can always redefine $F(\gamma,\hat \rho)$ to absorb the singular part of the density. Once this has been done, the matching calculation will determine the correct value of $F$ to ensure that physical observables are independent of which value we chose for $\alpha$.  As a check of this statement we have also performed these calculations for general $\alpha$.  The matching coefficient $F(1,\hat \rho)$ was modified, but $F'(1,\rho)$ was not. While the ``fluid'' density and pressure depended on $\alpha$, the computed fluid flows and \emph{total} energy-momentum tensor were ultimately independent of $\alpha$, as claimed.

\subsection{How the EFT Reproduces d'Alembert's Paradox}
As previously mentioned it is a text book calculation to show that the net force in a stationary potential flow vanishes \cite{landau1959fm}.
This is known as d'Alembert's paradox , which is  the counter-intuitive result that the drag force is zero when the fluid is moving at fixed velocity with respect to the
particle. This result typically follows by solving the boundary value problem, and integrating the pressure around the particle.
But in the EFT we have dispensed with the boundary value problem. Indeed if we took the full theory solution to the flow
and plugged into the RHS of (\ref{eom}) we would not find zero. In fact that process would be ill defined as in the EFT we have already coarse grained over scales of order the radius.  That is, the details of the flow at the edges have been integrated out.
The question arises as to what we take for the velocity gradient at the position of the particle?
The proper way to understand this issue follows from considering the meaning of the Euler equation in the EFT where the
sphere is now acting like a vorticial point source that has an effective charge proportional to the relative velocity between
the particle and the fluid. We may gain insight into this issue by asking the analogous question in the context of  a massive particle generating
a gravitational field.  In that case the field is infinite at the source, and we know how that this corresponds to an infinite
renormalization which we may set to zero. That is the field at the position of the particle, {\it generated by the particle}
is taken to be zero.
This is analogous with how the self force is implemented in the EFT approach to the self force problem \cite{Cheung:2024byb}.
This does not mean that there is {\it never} a force on the point particle. If we had some other external
sources  (point particle in the EFT) then the field generated by that source will generate a non-zero gradient for the other
particle.

 Note that even though we fixed the matching coefficients in a non-relativistic context, once we have fixed $F$ and $F^\prime$ we can return to the fully invariant EFT.  This is the power of the EFT, we can fix the matching coefficient 
in the simplest of states, since it is a universal short distance coefficient. This action is valid for any laminar flow despite the fact that
we calculate the matching coefficient in a potential flow. This may seem like an  insignificant accomplishment given that to the order we are working
the incompressible and irrotational nature of the flow is sub-leading. The question then arises whether or not for potential flow (which is solvable)
one could deduce the equations of motion by solving for the force on the sphere and then simply treating the sphere as a particle, thus avoiding the need to use the EFT in the first place? The answer is no for the following reasons.
It is important to recall that the functions $F(\rho)$ and $F^\prime(\rho)$ should be evaluated on the surface of the sphere when matching, which
by assumption, {\it would} take the same value  when extrapolated to the origin, in the point particle picture. The errors incurred
in this assumption being subleading in the gradient expansion.  We may then match to the case of an incompressible
flow. However, once we have determined how the functions $F,F^\prime$ depend upon $\rho$ we may relax the assumption of
incompressibility, as $\rho$ is free to vary over distances much larger than the radius of the sphere. If we we're interested
in higher-order derivative corrections this would entail new matching coefficients.

Thus the EFT has allowed
us to lift the force law determined by the potential flow to a non-potential flow. So that in the equations of motion for the incompressible
case will differ, in a calculable way 
 \beq
M\frac{d  v^i}{dt}=\frac{d}{dt}(\rho V_p   u^i -\frac{1}{2}\rho V_p( v - u)^i).
 \eeq

\subsection{The Relativistic Modified Euler Equations}
Our expansion remains valid in cases of fully relativistic velocities, provided the relative velocity between the fluid and the compact object remains small to $c$. This allows for a straightforward modification of the relativistic Euler equation within the same framework.
Once again, Conservation of the relativistic energy-momentum tensor allows us to derive the relativistic modified Euler and continuity equations.
\begin{equation}
    \partial_{\mu} T_{fl}^{\mu \nu} = u^{\mu} u^{\nu} \partial_{\mu} (\rho + p ) + (\rho + p )   u^{\mu} \partial_{\mu} u^{\nu}
+  (\rho + p )   u^{\nu} \partial_{\mu} u^{\mu} - g^{\mu \nu} \partial_{\mu} p .
\end{equation}
The projector to the plane perpendicular to the fluid velocity is given by
\begin{equation}
    h^{\alpha}_{\beta} = \delta^{\alpha}_{\beta} - u^{\alpha}u_{\beta}, 
\end{equation}
which allows us to extract the relativistic Euler equation,
\begin{equation}
       h^{\alpha}_{\nu} \partial_{\mu} T_{fl}^{\mu \nu} =    (\rho + p )   u^{\mu}  \partial_{\mu} u^{\alpha} - g^{\mu \alpha} \partial_{\mu} p.
\end{equation}
Finally, conservation of the full energy-momentum tensor leads to the relativistic modified Euler equation,
\begin{equation}
    (\rho + p )   u^{\mu}  \partial_{\mu} u^{\alpha} - g^{\mu \alpha} \partial_{\mu} p 
    = h^{\alpha}_{\nu} \partial_{\mu} T_{pp}^{\mu \nu}.
\end{equation}
where $T_{pp}$ is given in (\ref{Tpp}).
While the relativistic modified continuity equation is obtained by projecting the conservation of the full energy-momentum tensor onto the fluid velocity.
\begin{equation}
\label{cont}
    u^{\mu} \partial_{\mu} \rho  +  (\rho + p )    \partial_{\mu} u^{\mu} = u_{\nu} \partial_{\mu} T_{pp}^{\mu \nu} 
\end{equation}

\subsection{Higher order corrections}

Let us consider generalizing the action by including derivative corrections. Any term we add will be multiplied by
some function of the leading order invariants $\frac{\dot x \cdot u}{\sqrt{\dot x^2}}$ and $\rho$.  At the one derivative level we have two possibilities is $\partial \cdot u$ and $(\partial_\mu \rho u^\mu)$. However we can use the equations of motion (\ref{cont})  
to eliminate one of those terms leaving
\beq
S_{1}= \int d\lambda \sqrt{\dot x^2}\l[(\partial \cdot u) G_1[\frac{\dot x \cdot u}{\sqrt{\dot x^2}},\rho]
\r],
\eeq

We can again expand $G_1$ around $\gamma=1$ before, but
 it is no longer possible to match using the analytic solution.  
We may match by calculating numerically.
The new equations of motion, in the non-relativisitic limit, which correct eq.(\ref{landau}) are given by
\beq 
M((1-\hat \rho)\dot v^i+F^\prime (\dot u^i- \dot v^i))=MF^\prime(\vec v-\vec u)\cdot\partial^i \vec u - \partial_{i}(\partial\cdot u)G_{1}[1,\rho]-(\partial\cdot u)\frac{\partial G_{1}[1,\rho]}{\partial \rho} \partial_{i}\rho\,.
\eeq
To our knowledge this result has not appeared in the literature.

 \subsection{The coupling to the metric in the Newtonian limit}

The action \eqref{eq:EFTaction} defining our EFT is  covariant, so describing the interaction with gravity is completely straightforward. In general one might be interested in using this effective theory to study environmental effects on relativistic gravitational dynamics, as can be relevant for astrophysical compact binary inspirals.  While such potential applications are interesting, in this work we will illustrate the utility of this EFT in just the simplest gravitational setting, Newtonian gravity. In doing so we will derive known results with relative ease.

For weak gravitational fields, $h_{\mu\nu}=g_{\mu\nu}-\eta_{\mu\nu} \ll 1$, the action simplifies to
\begin{equation}
     S=-M\int d\lambda\,\sqrt{\dot x^2}   F\l( \frac{\dot x \cdot u}{\sqrt{\dot x^2}}, \rho\r)+\frac{1}{2}\int d^{4}x \sqrt{-g}T^{\mu\nu}h_{\mu\nu}\,, 
\end{equation}
 where the inner products in the first term are now taken with the flat metric. In the non-relativistic limit $T^{00}$ dominates and this reduces to
  \begin{align}
 S=-M\int dt \l(F(1,\hat{\rho})(1-\frac{v^2}{2})+\frac{1}{2}F^\prime(1,\hat{\rho})(\vec v-\vec u)^{2}\r)+\frac{MF(1,\hat{\rho})}{2}\int dt\,h_{00}(x(t))\,.
 \end{align}

Let us investigate a few consequences of this. First, for simplicity, let's set $\vec{u}=0$.  In doing so it becomes clear that the gravitational and inertial masses differ
   \begin{align}
 S -M\int dt +\int dt \left(M(F-F')\frac{v^{2}}{2}+MF\frac{h_{00}(x(t))}{2}\right)\,.
 \end{align}
 If we insert the values for $F, F'$ determined by matching, and introduce the Newtonian potential $\Phi=-\frac{1}{2}h_{00}$,  the equation of motion is then
 \begin{equation}
     \frac{d^{2}x^{i}}{dt^{2}}=-\frac{\rho_{ob}-\mu\rho_{fl}}{\rho_{ob}+\frac{1}{2}\mu\rho_{fl}}\partial^{i}\,\Phi\,.
 \end{equation}
We see that the gravitational attraction of the particle towards an external source is lessened by the presence of the fluid.  In fact, if the fluid has the same mass density as the object then the object will not accelerate towards the gravitating body at all! We can check here the well known physics of a bubble in a uniform gravitational field $\Phi=gz,\,\rho_{ob}=0$,
\begin{equation}
    \frac{d^{2}z}{dt^{2}}=+2g\,
\end{equation}
seeing that indeed bubbles do rise, and they also accelerate at twice the standard gravitational acceleration of matter.

We are now equipped to understand what Archimedes was thinking some 2000+ years ago. Archimedes must have realized that the presence of a nonzero fluid density spontaneously breaks local Lorentz invariance so that below the symmetry breaking scale the ratio of gravitational to inertial masses is no longer protected from renormalization!

 Adding a little more complexity we will work towards a description of the gravitational dynamics of two bodies submerged in fluid.  To add a second body we would have two worldlines and two matching functions $F_{1,2}$. In the non-relativistic limit though,  since $T^{00}$ dominates we can readily compute the mutual gravitational potential by taking the known potential and substituting the masses for the relevant coefficients here, 
 \begin{equation}
     V_{N}(|x_{1}-x_{2}|)=-\frac{G(m_{1}-\mu\rho V_{p\,1})(m_{2}-\mu\rho V_{p\,2})}{|x_{1}-x_{2}|}\,.
 \end{equation}
Where we've assumed $\vec{u}=0$. Again, we see that buoyancy effects modify the mutual attraction. When both objects are more dense than the fluid, they attract.  When one object is more dense but the other is less then they repel (again, think Archimedes). Amusingly we've also derived a third result, that two bubbles in a fluid will gravitationally attract one another. Our formalism allows us to
reproduce these well known results in a way that allows us to calculate relativistic corrections in a straight forward fashion.

  \section{Allowing for viscosity}
  
  There are many ways to include viscosity in mechanical problems, but the closed time path formalism  (or in-in or Keldysh formulation) is probably the best formulated for the EFT. 
 It is more often utilized in out of equilibrium  systems, but  is also well suited for dissipative systems since it is time asymmetric.
  The reader interested in the details of the subject may consult \cite{Liu:2018kfw,Akyuz:2023lsm}.  The path and the degrees of freedom are doubled such that we can write the  world-line action as 
  \beq
  S= \int (L( {\dot x_+}, u_+,B_+)d\lambda_+-L( {\dot x_-}, u_-,B_-)d\lambda_-)
  \eeq
  The action should be invariant under two separate
  world-line reparameterization invariances
  $Diff_1 \otimes Diff_1$.
  It is convenient to work in the Keldysh basis  
  \beq
   x^\mu_a=  x^\mu_+- x^\mu_- ~~~~~~~~ x_r^\mu=\frac{1}{2}( x^\mu_++ x^\mu_-),
  \eeq
  with similar definitions implied for the fluid variables. The equations of motion are then determined by
  varying the action with respect to the $x_a$ and then setting  it to zero, so that, classically at least, we only need to consider terms in the
  action  linear $x_a$. To generate dissipation we add terms to the action which
  involve both types of variables\footnote{This would not be the case if we were interested in quantum effects. The general principles for writing down the action are discussed in \cite{Liu:2018kfw}. Moreover, in this work we are truncating to only the lowest order dissipative terms and not explicitly writing the terms which lead to fluctuations. To be consistent with the thermodynamics of the underlying microscopic theory, one should also include such fluctuation terms with coefficients related to the coefficients of the dissipation terms via ``dynamical KMS symmetry'', however these can be suppressed at low temperatures. }  We must do so in a way
  which is consistent with our symmetry group
  \beq
  G = Diff_1 \otimes Diff_1 \otimes Diff_3 \otimes P_4,
  \eeq
  where $P_4$ is the 4-D Poincare group and $Diff_3$ are the group of volume preserving diffeomorphisms acting on the fluid elements.

  If we try covariantize the action we are immediately struck by the fact that $x_{a,r}$  are not tensorial objects.  Moreover, we have two different reparameterization groups for $x_-$ and $x_+$.
While the action need not depend upon $x_r$, it will have to depend upon $x_a$ or some covariantization thereof.
Intuitively we can guess the relevant tensorial generalization of $x_a$ should be given that it needs to  reduces to
the $x_+- x_-$ in the flat space limit. Thus it  should be formed from $\partial_\mu \sigma$ where
sigma is the squared geodesic distance between $x_+$ and $x_-$. As we show in the appendix this is indeed correct.

There  exists a  methodology for building tensorial objects from two points (recall that $x_-=x_1-x_2$) for a review see \cite{Poisson:2011nht}. 
First defining  $x_r$ as being the point along the geodesic (parameterized by $s$) $z^\mu(\lambda,s)$ halfway  between $x_+$ and $x_-$
\begin{equation}
    x^{\mu}_{r}(\lambda)=z^{\mu}(\lambda,s=\sqrt{\frac{\sigma}{2}}),
\end{equation}
where $\sigma(x_-,x_+)$ is related to the geodesic distance bewteen $x_-$ and $x_+$, 
while
\begin{equation}
    x_{a\,\mu}=\frac{\partial}{\partial x_{r}^{\mu}}\sigma(x_{r},x_{+})\,.
\end{equation}
$x_r$ is invariant under the off-diagonal combination of
the two RPIs, whose effect is to change the geodesic curve passing through $x_r$. We may use this latter symmetry to impose that the tangent of
the geodesic is orthogonal to $\dot x_r$. 
We can then form tensors which are invariant under residual  RPI
\begin{equation}
    \frac{\dot x_{r}^{\mu}}{\sqrt{\dot x_{r}^{2}}} \qquad \textrm{and}\qquad  \Pi_{\mu \nu}x_a^\mu\equiv X_{a\,\mu}\,,
\end{equation}
where $\Pi_{\mu \nu}$ is transverse projector
\beq
\Pi_{\mu \nu} \equiv \left(g^{\mu\nu}-\frac{\dot x_{r}^{\mu}\dot x_{r}^{\nu}}{\dot x_{r}^{2}}\right).
\eeq

   At lowest order in derivatives  the unique interaction Lagrangian invariant under the  symmetries is 
 \beq
 S_{int}= \int \frac{u_r \cdot X_{a}}{\sqrt{\dot x_r}^2}  K\l( \frac{\dot x_r \cdot u_r}{\sqrt{\dot x_r^2}},\rho_r\r)d\lambda.
 \eeq
 To get the equations of motion we vary with respect to $x_a$ and then set 
$x_+=x_-$,
\beq
m\ddot x^\mu= \Pi^{\mu \nu} u_\nu \,K\l( \frac{\dot x_r \cdot u_r}{\sqrt{\dot x_r^2}},\rho\r)
\eeq

As in the conservative case we now expand the function $K$ in the small relative velocities
 and choose the global time as the affine parameter. At next to leading order in velocities we have
 \beq
 \boxed{
m\ddot x^i= \Pi^{\mu \nu} u_\nu (K\l( 1,\rho\r)+(K^\prime\l( 1,\rho\r)(\frac{\dot x_r \cdot u_r}{\sqrt{\dot x_r^2}}-1)+O((\gamma -1)^2)}
\eeq
This is the covariant generalization of the Stokes equation. As far as we are aware it has not appeared in the literature before. It was anticipated by \cite{Correia:2022gcs} who, rightly, commented that for generic configurations an extremely high viscosity would be required to keep the Reynold's number low while allowing for large velocities.

Now we would like to fix the matching coefficients $K(1,\rho)$ and $K^\prime(1,\rho).$
As before we can match by working in the non-relativistic limit. The leading order equation of motion is 
\beq
m \ddot { \vec x}= K(1,\rho)(\vec u - \dot{ \vec x})
\eeq

 The well known full theory solution for the force on a sphere in the Stokes limit (small Reynolds number) is given by
 $\vec F= 6 \pi \rho \nu (\vec u -\dot{\vec x})$, where $\nu$ is the kinematic viscosity. Thus we find $K(1,\rho)=6 \pi \rho \nu$. To match $K^\prime(1,\rho)$ we would need to calculate the corrections to Stokes law in the full theory. This can always be done 
 numerically.

  \section{Applications }

A primary motivation of this paper is to build a theory which allows one to calculate the corrections to gravitational wave signals from binary inspirals when the constituents are immersed in a fluid. At the outset we have assumed that the gradients in the fluid are small, i.e. that the flow is laminar, i.e small Reynolds number $Re < 2000$. This is not within the regime we would expect for an inspiral. The Reynolds number will scale like
$Re= \frac{R v}{\nu}$. The viscosity ($\nu$) of a fluid scales as $\nu \sim l v_T$, where $v_T$ is the thermal velocity and $l$ the mean free path.
Thus to get a sufficiently small Reynolds number we would need the fluid to be extremely hot and dilute, which would imply a small density and therefore any drag force would be negligible. In general we would expect
dynamical friction to be the dominant dissipative force during the inspiral. 

Furthermore, we have matched with hard wall boundary conditions which surely are inappropriate for neutron stars where we expect mass accumulation.
Of course for black holes the situation simplifies even further having purely absorptive boundary conditions. Indeed, once we can neglect drag forces on the object we no longer need to restrict ourselves to laminar flow, as the forces on the object will be purely gravitational. The only non-gravitational effect of the fluid particle interaction will be due to mass accretion, the net effect of which will of course be dependent upon the choice of the parameters. Given that we are still working in the point particle approximation
we may ask if in the turbulent regime we will still be able to capture that effect in a systematic fashion? We would expect that  a suitable averaging
of the fluid flux in a region of order $r$ around the particle should be sufficient but further work in 
this direction is needed.

Another interesting application would be to calculate induced forces between submersed objects beyond leading order buoyancy forces. In particular, we have in mind calculating the forces between objects due to bulk non-linearities.

\section{Discussion and Future Directions}

In this paper we have developed a point particle EFT formalism which generalizes the gravitational EFT \cite{Goldberger:2006bd} to allow for the coupling of compact objects to fluids for the purposes of reducing the complexity of the coupled Navier Stokes equation.
By considering the long wavelength approximation the EFT is capable of describing the motion of a compact object
in a fluid as a point particle.
There is no need to deal with nettlesome boundary conditions at each step. 
The physics of the boundary is encapsulated by 
some new point like interactions which arise 
in the equations of motion. The strength of these interactions is controlled by a set of constants which can be fixed  by performing a matching calculation.  

If we wish to calculate the forces on the object due to drag the theory is only valid when the flow is laminar, so that the gradients in the fluid density are small compared to the inverse size of
the objects. In this case we were able to write down a fully relativistic equation of motion and
fixed the coefficients up to fourth order in the
velocity difference in the perfect fluid case and to second order in the viscous case. In all cases our results were written completely covariant allowing for the propagation in a general space-time. 
The results in this paper can be generalized 
to include finite size effects, or spin  \cite{Porto:2005ac}. 

Given that the theory is only valid for laminar flow we don't expect it to be relevant for generic binary inspirals. However, for such cases, we expect drag forces to be negligible, with gravitational forces instead dominating. In this case there is no need to restrict ourselves to laminar flow anymore. The dominant effect of interactions between the object and the fluid
will be mass accretion which we believe can be treated statistically in a turbulent flow  despite the large gradients. At the level of the action dissipation is accomodated by adding new world-line degrees of freedon \cite{Goldberger:2005cd,Goldberger:2020wbx} Furthermore, since the theory is completely covariant we can calculate post-Newtonian corrections  systematically including the
interactions stemming from integrating over the fluid density.

\appendix
\section{Generally Covariant Keldysh Variables}

\subsection{Motivation and Results}

To describe dissipation in a Lagrangian formalism one typically introduces the in-in framework. In this framework the dynamical variables, say $(x, \phi)$, are doubled into two copies, and the dynamics of a standard conservative system is written as
\bea
\mathbf{S}[x_{+},x_{-},\phi_{+},\phi_{-}]=S[x_{+},\phi_{+}]-S[x_{-},\phi_{-}]\,.
\eea
The conservative nature of the underlying physics dictates the separation of the action into these two independent terms, ie. there is no coupling between plus and minus variables.  

If the $x^{\mu}$ represent some system of interest and the $\phi$ are a collection of environmental variables, one can integrate out the $\phi$ variables to obtain an in-in effective action for the $x^{\mu}$ variables~\cite{Schwinger:1960qe,Keldysh:1964ud,Feynman:1963fq}.  Quite generally the coupling to the environment will involve an exchange of conserved quantities such as energy-momentum, and the effective action for the system will reflect that the system itself no longer conserves energy-momentum (or in a quantum theory, the evolution may no longer be unitarity). The signature that the system is effectively ``open'' is that the effective action will involve coupling between the plus and minus variables, e.g.,
\bea
\mathbf{S}^{eff}[x_{+},x_{-}]=S'[x_{+}]-S'[x_{-}] + \Phi[x_{+},x_{-}]\,,
\eea
where $\Phi[x_{+},x_{-}]$ need not separate into plus and minus contributions. In practice, the underlying dynamics may be known and the effective theory can be explicitly computed by exactly integrating out the environment. When the exact dynamics are too complicated, or even unknown, one can resort to a ``bottom-up'' construction of $\Phi$ as an effective theory. For details see \cite{Liu:2018kfw}. 

Suppose the system of interest is a point particle moving through flat spacetime. One typically utilizes the in-in action by defining Keldysh variables
\bea
x^{\mu}_{r}=\frac{1}{2}(x_{+}^{\mu}+x_{-}^{\mu})\qquad x^{\mu}_{a}=x^{\mu}_{+}-x^{\mu}_{-}\,.
\eea
The main convenience provided by these variables is that when we expand the action for small $x_{a}$ we generate the standard classical equations of motion for $x_{r}$.  For example, when $\Phi=0$ we have
\bea\label{eq:flatspacekeldyshaction}
\mathbf{S}^{eff}[x_{r},x_{a}]=S'[x_{r}]-S'[x_{r}]+\int x_{a} \frac{\delta S'[x_{r}]}{\delta x_{r}}+\mathcal{O}(x_{a})^{2}\,.
\eea
In a limit where higher order terms in $x_{a}$ can be neglected, we see that extremizing the action with respect to $x_{a}$ then imposes the classical equation of motion for $x_{r}$.  If $\Phi$ is non-zero and contains terms linear in $x_{a}$, we can then get modified equations of motion for $x_{r}$ which are, in general, dissipative. Retaining high-order terms in $x_{a}$ leads to effective stochastic forces in the equation of motion for $x_{r}$.

In this paper we are interested in a generally covariant description so that we may describe a particle experiencing dissipation while moving in curved spacetime. There is no obstruction to introducing plus and minus variables, but there is an immediate confusion about how to implement Keldysh variables. For example the time derivatives $\tfrac{dx^{\mu}_{+}}{d\lambda}$ and $\tfrac{dx^{\mu}_{-}}{d\lambda}$ are vectors living in entirely different tangent spaces, so how can we add and subtract them?  In what follows we will provide a geometric construction which allows for a definition of Keldysh variables in curved spacetime. The crucial property possessed by the the flat space expression \eqref{eq:flatspacekeldyshaction}, which we will require of our curved space construction, is that when the effective action is expanded in powers of $x_{a}$, the terms independent of $x_{a}$ vanish and the remaining terms are local, i.e. given by the integral of an effective Lagrangian on the $x_{r}$ worldline.

The dynamical variables are $x_{\pm}(\lambda)$, and the underlying action has independent reparameterization invariance under $\lambda \rightarrow \tilde{\lambda}(\lambda)$ for each worldline. To build our Keldysh variables we need a geometric relationship between points on the two worldlines. Due to the two-fold reparameterization invariance, for a given parameterization of $x_{-}$ we have an infinite set of physically equivalent choices for mapping a given point $x_{-}(\lambda)$ to a point on the $x_{+}$ worldline. To proceed we will assume that the worldlines are in each other's normal convex neighborhood. That is, for every pair of points on $x_{-}$ and $x_{+}$ there is a unique geodesic linking the two.\footnote{We expect a sufficient condition for our construction to be that not all points on the two worldlines are in each other's normal convex neighborhood, but just that there exist collections of open sets, $\mathcal{U}_{\pm}=\{U_{\pm}\}$, which cover each worldline and that for every element $U_{+}$ there exists an element $U_{-}$ in its normal convex neighborhood, and that there are sufficient consistency conditions on the overlaps of all these sets. }  

Concretely, what we require is that for a given parameterization $x_{-}(\lambda)$ there exists a family of geodesics which are indexed by $\lambda$ and parametrized by $s$, $z^{\mu}(\lambda,s)$, such that $z^{\mu}(\lambda, s_{0}) = x^{\mu}_{-}(\lambda)$ and that $z^{\mu}(\lambda,s_{1})$ is a unique point on $x_{+}$ for all $\lambda$. For a choice of geodesic family $z^{\mu}(\lambda,s)$  and a parameterization $x_{-}(\lambda)$ we then have an induced parameterization $x_{+}(\lambda)$. The two-fold reparameterization invariance of the dynamics will manifest itself in a reparameterization invariance of $z^{\mu}$ in addition to invariance under the choice of geodesic family.  For a given point $x_{-}(\lambda)$ we could have considered any geodesic $z^{\mu}(\lambda,s)$ which connects to a point on $x_{+}$.

Once a geodesic family has been chosen, we can define Synge's worldfunction $\sigma(x_{-},x_{+})$.  This object is a biscalar on each wordline, and it's derivatives generate objects which are bitensors (see \cite{Poisson:2011nht} for details).  This is the object which allows us to link the points along each worldline in an invariant manner to yield tensorial quantities living in the same tangent space.  The definition of Synge's worldfunction is
\beq
\sigma(x_{-}(\lambda),x_{+}(\lambda))= \frac{1}{2} (s_{1}-s_{0})\int_{s_{0}}^{s_1}ds\, g_{\mu \nu}(z^\mu(\lambda,s)) \frac{dz^{\mu}(\lambda,s)}{ds}\frac{dz^{\nu}(\lambda,s)}{ds}\,,
\eeq
which is invariant under affine reparameterizations of $s$, and equal to one-half the squared geodesic distance between $x_{-}(\lambda)$ and the corresponding point  $x_{+}(\lambda)$. 
The quantity  \newline $\epsilon \equiv g_{\mu \nu}(z^\mu(\lambda,s)) \frac{dz^{\mu}(\lambda,s)}{ds}\frac{dz^{\nu}(\lambda,s)}{ds}$ is a constant as a consequence of the geodesic equation, and therefore numerically
\beq\sigma(x_{-}(\lambda),x_{+}(\lambda))= \frac{1}{2} (s_{1}-s_{0})^2\epsilon.
\eeq
For time/space-like geodesics we can choose $s$
to be the proper time/length and $\epsilon=\pm 1$.

Derivatives of this function are tangent to the geodesics
\begin{equation}\label{eq:syngederivative}
    \frac{\partial}{\partial x^{\mu}_{-}}\sigma(x_{-}(\lambda),x_{+}(\lambda)) = -(s_{1}-s_{0}) g_{\mu\nu}(x_{-}(\lambda))\frac{dz^{\mu}(\lambda,s_{0})}{ds}\,.
\end{equation}
An important property, which we will later use, is that this vector squares to the worldfunction
\begin{equation}
    g^{\mu\nu}(x_{-})\partial_{\mu}\sigma(x_{-},x_{+})\partial_{\nu}\sigma(x_{-},x_{+}) = 2\sigma(x_{-},x_{+})\,.
\end{equation}

This construction allows us to introduce Keldysh variables. Let us \emph{define} $x_{r}^{\mu}$ to be the point along the curve $z^{\mu}(\lambda,s)$ which, for each $\lambda$, is $1/2$ the geodesic distance from $x_{-}$ to $x_{+}$.  Said differently, if $s$ is taken as the proper length parameter (for space like geodesics) 
\begin{equation}
    x^{\mu}_{r}(\lambda)=z^{\mu}(\lambda,\sqrt{\frac{\sigma}{2}})\,.
\end{equation}

While we started from a description with $x_{\pm}$ as independent variables, we can now make the choice to use $x_{r}$ and $x_{+}$ as independent variables.  Thus $\partial_{\mu}(x_{r},x_{+})$ may be taken as a vector living along the worldline $x_{r}$ which is an independent variable to the curve $x_{r}$.  We then \emph{define} the Keldysh $x_{a}$ variable to be 
\begin{equation}
    x_{a\,\mu}=\frac{\partial}{\partial x_{r}^{\mu}}\sigma(x_{r},x_{+})\,.
\end{equation}
Whereas $x_{r}^{\mu}$ is not a vector, just a coordinate, we see that $x_{a\,\mu}$ is indeed a (co)vector.  Using the definition of the worldfunction it is easy to check that
\begin{equation}
    g^{\mu\nu}x_{a\,\mu}x_{a\,\nu}=8\sigma(x_{r},x_{+})=2\sigma(x_{-},x_{+})
\end{equation}
which is precisely the squared geodesic distance between $x_{-}$ and $x_{+}$, in agreement with the standard flat spacetime definition.

As alluded to previously there is a diagonal action of $Diff_{1}\otimes  Diff_{1}$ which corresponds to reparametrizations of $x_{r}(\lambda)$.  The off-diagonal group action however will leave $x_{r}$ invariant and will instead change the  curve $z^{\mu}(\lambda,s)$ passing through a given point $x_{r}(\lambda)$. With all of this redundancy in the choice of $z^{\mu}$, we are free to chose $z^{\mu}(\lambda,s)$ to intersect $x^{\mu}(r,\lambda)$ orthogonally, for each $\lambda$.  That is, we are free to choose $\partial_{\mu}\sigma$ such that $\dot{x}_{r}^{\mu}\partial_{\mu}\sigma=0$. We can manifest this orthogonality by building the action out of the building blocks
\begin{equation}
    \frac{\dot x_{r}^{\mu}}{\sqrt{\dot x_{r}^{2}}} \qquad \textrm{and}\qquad X_a^\mu\equiv\left(g^{\mu\nu}-\frac{\dot x_{r}^{\mu}\dot x_{r}^{\nu}}{\dot x_{r}^{2}}\right)x_{a\,\mu}\,,
\end{equation}
While the former is intuitive, the validity of the latter can be confirmed by noting that this projector is essential to make sure that the four-forces in the equation of motion for $x^{\mu}_{r}$ preserve the on-shell condition \begin{equation}
    0=\frac{d}{d\lambda}\dot{x}_{r}^{2}\propto 2\dot{x}_{r}^{\mu}\frac{\partial \mathcal{L}}{\partial x_{a}^{\mu}}\,,
\end{equation}
as required by RPI invariance.

\subsection{Further Geometric Intuition}

To make this discussion more self-contained, here we provide some more geometric details and intuition to explain why Synge's worldfunction is the natural object to use in this construction.

\begin{figure}
    \centering
    \includegraphics[width=0.5\linewidth]{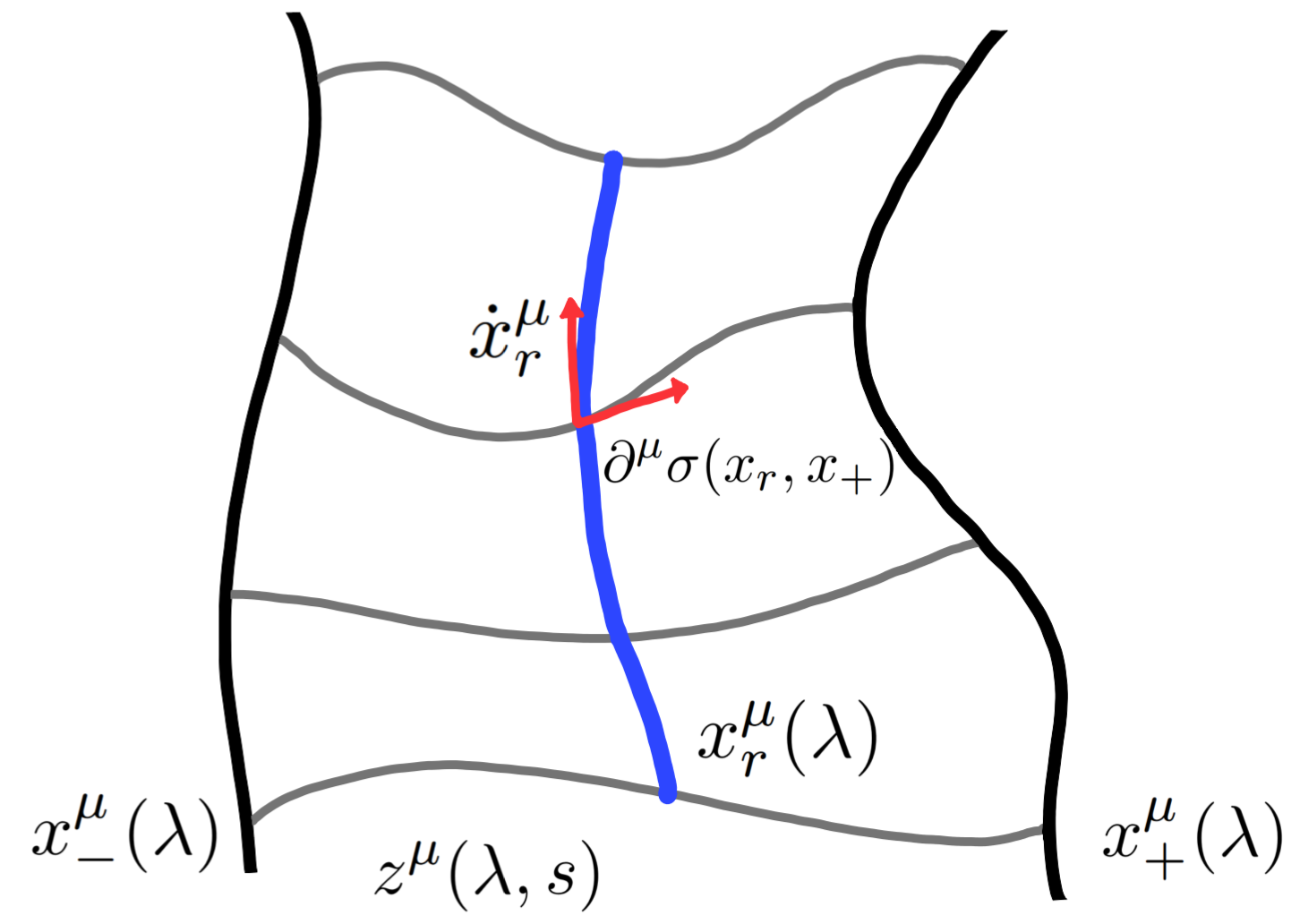}
    \caption{An illustration of our generally covariant Keldysh variable construction. A point on the left curve is labeled $x_{-}^{\mu}(\lambda)$, and is connected to a point on the right curve via a geodesic $z^{\mu}(\lambda,s)$. The intersection point on the right curve is then labeled $x_{+}(\lambda)$. The curve bisecting the geodesic family $z^{\mu}(\lambda,s)$ along the midline of proper length is called $x^{\mu}_{r}(\lambda)$. For a generic choice of geodesic family the tangent vectors $\dot{x}^{\mu}_{r}$  and $\partial^{\mu}\sigma$ at the point $x^{\mu}_{r}(\lambda)$ are not necessarily orthogonal, but can be made so by a suitable choice of geodesics $z^{\mu}(\lambda,s)$.}
    \label{fig:geodesicKeldysh}
\end{figure}

We start with the $\pm$ variables, for which the action is given by integrals of functions along each of the $x_{\pm}$ worldlines. We desire a curved space generalization of Keldysh variables which will involve an action integrated only over some $x_{r}$ worldline, with some tensorial quantity $x_{a}^{\mu}$ living on the $x_{r}$ worldline.  This Keldysh effective action, even in the absence of dissipative and stochastic contributions from $\Phi[x_{+},x_{-}]$, must be able to describe the Langranians $L(x_{\pm})$ which live on the $x_{\pm}$ worldlines. The question is then, how can functions at $x_{\pm}$ be described using objects living on $x_{r}$. To answer this, we employ the geodesic family $z^{\mu}(\lambda, s)$.

Once a geodesic family has been chosen there is a way to describe functions on $x_{+}$ via functions on $x_{-}$.  For an arbitrary smooth function $f(x)$ we have that
\begin{align}
f(x_{+}(\lambda))=f(z(\lambda,s_1 ))&=\sum_{n}\frac{(s_{1}-s_{0})^{n}}{n!}\frac{d^{n}}{ds^{n}}f(z(\lambda,s))\bigg|_{s=s_0 }.
\end{align}
If $z^{\mu}$ is an affine parameterized geodesic then the higher $s$ derivatives take a very simple form
\begin{align}
f(x_{+}(\lambda))&=\sum_{n}\frac{(s_{1}-s_{0})^{n}}{n!}\frac{dz^{\mu_{1}}(\lambda,s_{0})}{ds}\cdots \frac{dz^{\mu_{n}}(\lambda,s_{0})}{ds} \nabla_{\mu_{1}}\cdots \nabla_{\mu_{n}}f(x_{-}(\lambda))\bigg|_{s=s_{0}} \nonumber \\
&=\exp\left((s_{1}-s_{0}) \frac{dz^{\mu}(\lambda,s_{0})}{ds}\nabla_{\mu}\right)f(x_{-}(\lambda)).
\end{align}
Thus, the function $f(x_{+}(\lambda))$ can be computed using vectors in the tangent space of $x_{-}$. In principle any affine parameter $s$ can be used, but the above has a nice $s$-independent description in terms of the derivative of Synge's worldfunction \eqref{eq:syngederivative},
\begin{equation}
    f(x_{+})=\exp\bigg(-g^{\mu\nu}(x_{-})\partial_{\mu}\sigma(x_{-},x_{+})\nabla_{\mu}\bigg)f(x_{-})\,.
\end{equation}
where we've suppressed the $\lambda$ dependence in each function, and the partial derivative is taken with respect to the first argument.

Similarly, once the mid-point worldline $x_r$ has been introduced we can also use the derivatives of Synge's worldfunction to describe functions at both $x_{-}$ and $x_{+}$,
\begin{align}
f(x_{+})&=\exp\bigg(-g^{\mu\nu}(x_{r})\frac{\partial}{\partial x_{r}^{\mu}}\sigma(x_{r},x_{+})\nabla_{\mu}\bigg)f(x_{r})\,, \nn \\
f(x_{-})&=\exp\bigg(-g^{\mu\nu}(x_{r})\frac{\partial}{\partial x_{r}^{\mu}}\sigma(x_{r},x_{-})\nabla_{\mu}\bigg)f(x_{r})
\end{align}
Since we've chosen $x_{r}$ to be equidistant from both of these curves we have that
\begin{equation}
    \frac{\partial}{\partial x_{r}^{\mu}}\sigma(x_{r},x_{+})=-\frac{\partial}{\partial x_{r}^{\mu}}\sigma(x_{r},x_{-})\,,
\end{equation}
so that the expressions can be unified as
\begin{align}
f(x_{\pm})&=\exp\bigg(\mp g^{\mu\nu}(x_{r})\frac{\partial}{\partial x_{r}^{\mu}}\sigma(x_{r},x_{+})\nabla_{\mu}\bigg)f(x_{r})\,, 
\end{align}
Finally, we can define $x_{a\,\mu}\equiv\partial_{r\,\mu}\sigma(x_{r},x_{+})$. We can chose, instead of $x_{\pm}$, to take $x_{a\,\mu}$ and $x_{r}^{\mu}$ as the independent dynamical variables, w–ith $x_{r}^{\mu}$ being the coordinates of a worldline and $x_{a\,\mu}$ being a tensor on the $x_{r}$ worldline. In terms of these variables we clearly have the ability to describe functions along the $x_{\pm}$ worldlines via
\begin{align}
f(x_{\pm})&=\exp\bigg(\mp x_{a}^{\mu}\nabla_{\mu}\bigg)f(x_{r})\,.
\end{align}
We have thus derived purely geometric quantities which posseses the desired properties and reduce to the familiar Keldysh variables in the flat space limit.

\bibliography{main.bib}

\end{document}